# Super-resolution far-ultraviolet equivalent-wavelength interferometry combining the phase of several visible-wavelength interferograms

**MANUEL SERVIN,*  MOISES PADILLA, AND GUILLERMO GARNICA**

[1]*Centro de Investigaciones en Optica A. C., 37000 Leon Guanajuato, Mexico.*
*mservin@cio.mx

**Abstract:** A far-ultraviolet (FUV) equivalent-wavelength super-resolution interferometric technique is proposed. This FUV equivalent-wavelength interferometric method combines four demodulated phases from four temporal-sets of visible interferograms. Here FUV super-resolution interferometry is defined as the estimation of a modulating phase coming from an FUV equivalent-wavelength illumination laser. To this end we need to combine the demodulated phase of four visible-wavelength interferograms. FUV equivalent-wavelength phase-sensitivity is of course beyond the phase-information capacity of a single visible-wavelength interferogram. To break this visible-wavelength barrier we use the phase-information provided by four or more interferograms in the visible range. Having the four demodulated phases we calculate a phase-difference and the sum of the four phases which is the FUV-equivalent super-resolution phase. The phase-difference is in the infrared phase-sensitivity range and it is assumed non-wrapped. On the other hand the phase-sum is in the FUV phase-sensitivity range and it is highly-wrapped. As shown herein it is possible to unwrap the phase-sum in the temporal domain using the phase-difference and our previously reported extended-range 2-step temporal phase-unwrapper. Of course higher than FUV equivalent phase-sensitivity interferometry may be obtained by increasing the number of independent estimated phases from visible-wavelength interferograms. As far as we know, this FUV equivalent-wavelength super-resolution interferometric technique has the highest phase-sensitivity and highest signal-to-noise ratio ever reported to this date.

---

## 1. Introduction

Temporal phase-shifting interferometry is well known and has been widely used for many years in optical metrology [1,2]. Wyant was the first researcher to use two close laser-wavelengths ($\lambda_1 \approx \lambda_2$) to test an optical aspheric surface with infrared equivalent sensitivity wavelength of [3],

$$\lambda_{eq} = \frac{\lambda_1 \lambda_2}{|\lambda_1 - \lambda_2|} \tag{1}$$

Given that $\lambda_1 \approx \lambda_2$, the equivalent wavelength $\lambda_{eq}$ may easily be in the infrared frequency range i.e. $\lambda_{eq} >> \{\lambda_1, \lambda_2\}$. The infrared-equivalent wavelength $\lambda_{eq}$ is obtained by the following phase-difference [3-12],

$$\frac{2\pi}{\lambda_1}W(x,y) - \frac{2\pi}{\lambda_2}W(x,y) = \frac{2\pi}{\lambda_{eq}}W(x,y) \tag{2}$$

Being $W(x,y)$ the optical wavefront. Double-wavelength infrared equivalent-sensitivity interferometry was improved by Polhemus [4] and later on by Cheng [5,6] using digital phase-shifting phase-demodulation. Afterwards Onodera et al. used Fourier phase-demodulation methods for profiling structures with equivalent lambda depth size [7]. Unfortunately, the demodulated phase was over-smoothed due to the over filtering of too close Fourier diffraction orders [7]. This in turn was followed by a large number of double-wavelength Fourier and phase-shifting phase-demodulation methods in such diverse applications as double-wave holographic microscopy ]8], extended range optical metrology [9], two-step digital holography [10], multi-wavelength extended-range contouring [11], and two-wavelength surface profiling [12]. Even though these methods [3-12] were applied to a variety of different experimental situations, all of them share the same mathematical background which may succinctly stated as follows: given two laser-interferometric optical phases $\varphi_1(x,y) = (2\pi/\lambda_1)W(x,y)$ and $\varphi_2(x,y) = (2\pi/\lambda_2)W(x,y)$, calculate their phase difference $\varphi_D(x,y) = \varphi_1(x,y) - \varphi_2(x,y)$ such that $\varphi_D(x,y)$ is not wrapped i.e. $\varphi_D(x,y) \in (0, 2\pi)$. In this way researches rightly claim having high-depth infrared-equivalent phase-range measuring capabilities and requiring no phase-unwrapping.

Here we propose to combine four phase-shifted demodulated phases $\varphi_1(x,y)$, $\varphi_2(x,y)$, $\varphi_3(x,y)$ and $\varphi_4(x,y)$ obtained from four sets of temporal phase-shifted visual-wavelength interferograms. As we demonstrate, the phase-sensitivity of the sum $\varphi_1(x,y) + \varphi_2(x,y) + \varphi_3(x,y) + \varphi_4(x,y)$ may go up to the FUV phase-sensitivity range. The presentation of this paper is as follows: in section 2 we present the four visible-wavelength phase estimations to obtain FUV super-resolution interferometry. In section 3 we compute the phase-difference and the phase-sum needed to mathematically construct our FUV equivalent phase-sensitivity interferometer. In section 4 we show how temporally unwrap the FUV-equivalent interferometric phase-sum, taking our infrared-equivalent phase-difference as first estimation. In section 5 we show the signal-to-noise power-ratio gain of the FUV-equivalent phase-sum with respect to the infrared-equivalent phase-difference and in section 6 we draw some conclusions.

## 2. Far-ultraviolet (FUV) interferometry using phase-difference and phase-sum

Temporal phase-shifting interferometry is a well known technique that has been used for many years [1,2]. The mathematical model for the interferograms $I_1(x,y,t)$, $I_2(x,y,t)$, $I_3(x,y,t)$ and $I_4(x,y,t)$ taken by the CCD camera are,

$$
\begin{aligned}
I_1(x,y,t) &= a(x,y) + b(x,y)\cos\left[\frac{2\pi}{\lambda_1}W(x,y) + n_1(x,y) + \frac{2\pi}{4}t\right]; \\
I_2(x,y,t) &= a(x,y) + b(x,y)\cos\left[\frac{2\pi}{\lambda_2}W(x,y) + n_2(x,y) + \frac{2\pi}{4}t\right]; \\
I_3(x,y,t) &= a(x,y) + b(x,y)\cos\left[\frac{2\pi}{\lambda_2}W(x,y) + n_3(x,y) + \frac{2\pi}{4}t\right]; \quad (\lambda_1 > \lambda_2); \\
I_4(x,y,t) &= a(x,y) + b(x,y)\cos\left[\frac{2\pi}{\lambda_2}W(x,y) + n_4(x,y) + \frac{2\pi}{4}t\right]; \quad t = \{0,1,2,3\}.
\end{aligned}
\quad (3)
$$

Where $a(x,y)$ is the background illumination; $b(x,y)$ the fringe contrast; $W(x,y)$ is the measuring optical wavefront and $\{n_1(x,y), n_2(x,y), n_3(x,y), n_4(x,y)\}$ are four independent samples of a Gaussian zero-mean stationary random process [15]. The four noise fields are assumed to have small amplitude *i.e.* $n_{1-4}(x,y) \ll 2\pi$. Finally $\{\lambda_1, \lambda_2\}$ are two different but close optical laser wavelengths ($\lambda_1 \approx \lambda_2$) [3-12].

Using the least-squares 4-step phase-shifting algorithm we demodulate the fringes in Eq. (3) obtaining the following four complex-valued analytic signals [2],

$$
\begin{aligned}
A_1(x,y)e^{i[\varphi_1(x,y)+n_1(x,y)]} &= \sum_{t=0}^{3} I_1(x,y,t)e^{i\left(\frac{2\pi}{4}\right)t}; \\
A_2(x,y)e^{i[\varphi_2(x,y)+n_2(x,y)]} &= \sum_{t=0}^{3} I_2(x,y,t)e^{i\left(\frac{2\pi}{4}\right)t}; \\
A_3(x,y)e^{i[\varphi_2(x,y)+n_3(x,y)]} &= \sum_{t=0}^{3} I_3(x,y,t)e^{i\left(\frac{2\pi}{4}\right)t}; \\
A_4(x,y)e^{i[\varphi_2(x,y)+n_4(x,y)]} &= \sum_{t=0}^{3} I_4(x,y,t)e^{i\left(\frac{2\pi}{4}\right)t}.
\end{aligned}
\quad (4)
$$

As Eq. (4) shows, the 3 phases $\varphi_2(x,y)$ are obtained from 3 independent measurements, therefore the noise fields $n_2(x,y)$, $n_3(x,y)$ and $n_4(x,y)$ are independent samples of the same stochastic process [15]. It is well known that 3 phase-shifted fringes is the minimum interferograms number, however increasing to 4 phase-shifted ($2\pi/4$ radians/interferogram) fringes one decreases substantially the harmonic content in the demodulated phase [2].

## 3. Calculating the phase-difference and phase-sum for FUV interferometry

Using the four analytic signals in Eq. (4) the following two products are computed,

$$A_1 e^{i(\varphi_1+n_1)} \left[ A_2 e^{i(\varphi_2+n_3)} \right]^* = A_1 A_2^* e^{i[\varphi_1-\varphi_2+n_1-n_2]},$$
$$A_1 e^{i(\varphi_1+n_1)} A_2 e^{i(\varphi_2+n_2)} A_3 e^{i(\varphi_2+n_3)} A_4 e^{i(\varphi_2+n_4)} = A_1 A_2 A_3 A_4 e^{i[\varphi_1+3\varphi_2+n_1+n_2+n_3+n_4]}. \tag{5}$$

Where $[x]^*$ stand for the complex-conjugate of $x$. The two wavelengths $\{\lambda_1, \lambda_2\}$ are chosen close enough so the phase difference,

$$\varphi_D(x,y) + n_D(x,y) = angle\left[ A_1 A_2^* e^{i[\varphi_1-\varphi_2+n_1-n_2]} \right], \tag{6}$$

is in the infrared equivalent-wavelength phase-sensitivity and non-wrapped i.e. $\varphi_D(x,y) \in (0, 2\pi)$. However the phase sum $\varphi_S^W(x,y)$,

$$\varphi_S^W(x,y) + n_S(x,y) = angle\left[ A_1 A_2 A_3 A_4 e^{i[\varphi_1+3\varphi_2+n_1+n_2+n_3+n_4]} \right], \tag{7}$$

is highly wrapped because $\varphi_1(x,y) + 3\varphi_2(x,y)$ is within the FUV equivalent phase-sensitivity range. The phase-difference and the phase-sum have equivalent illumination wavelengths of,

$$\frac{2\pi}{\lambda_1} W(x,y) - \frac{2\pi}{\lambda_2} W(x,y) = \frac{2\pi}{\lambda_D} W(x,y),$$
$$\frac{2\pi}{\lambda_1} W(x,y) + 3\left[ \frac{2\pi}{\lambda_2} W(x,y) \right] = \frac{2\pi}{\lambda_S} W(x,y). \tag{8}$$

From this equation one may derive the two equivalent wavelengths $\lambda_D$ and $\lambda_S$ as,

$$\lambda_D = \frac{\lambda_1 \lambda_2}{|\lambda_1 - \lambda_2|}; \quad \text{for} \quad \varphi_D(x,y) = \varphi_1(x,y) - \varphi_2(x,y),$$
$$\lambda_S = \frac{\lambda_1 \lambda_2}{3\lambda_1 + \lambda_2}; \quad \text{for} \quad \varphi_S(x,y) = \varphi_1(x,y) + 3\varphi_2(x,y). \tag{9}$$

If $\lambda_1 \approx \lambda_2$, then $\lambda_D$ may be in the infrared wavelength range, i.e. $\lambda_D \gg \{\lambda_1, \lambda_2\}$. In contrast $\lambda_S$ is in the FUV wavelength range, i.e. $\lambda_S \ll \{\lambda_1, \lambda_2\}$. In other words, $\varphi_D(x,y)$ have infrared sensitivity while $\varphi_S(x,y)$ have FUV phase-sensitivity. For example, assume two typical laser wavelengths of $\lambda_1 = 632.8$nm and $\lambda_2 = 532$nm. Then values for $\lambda_D$ and $\lambda_S$ are,

$$\lambda_D = 3,340 \text{ nm}; \quad \text{for} \quad \varphi_D(x,y) = \varphi_1(x,y) - \varphi_2(x,y),$$
$$\lambda_S = 138.5 \text{ nm}; \quad \text{for} \quad \varphi_S(x,y) = \varphi_1(x,y) + 3\varphi_2(x,y). \tag{10}$$

This is equivalent of using two illuminating laser wavelengths: one in the infrared with $\lambda_D = 3,340$ nm, and the other with FUV equivalent-wavelength $\lambda_S = 138.5$ nm.

Then the sensitivity gain $G$ between the infrared-equivalent $\varphi_D(x,y)$ and the FUV-equivalent $\varphi_S(x,y)$ is,

$$G = \frac{\|\varphi_S(x,y)\|}{\|\varphi_D(x,y)\|} = \frac{\lambda_D}{\lambda_S} = \frac{3\lambda_1 + \lambda_2}{|\lambda_1 - \lambda_2|}, \tag{11}$$

Where $\|\cdot\|$ stand for the maximum norm. This means that the phase-sensitivity gain $G$ between $\varphi_D(x,y)$ and unwrapped $\varphi_S(x,y)$ is,

$$\varphi_S(x,y) = \frac{\lambda_D}{\lambda_S}\varphi_D(x,y) = \frac{3\lambda_1 + \lambda_2}{|\lambda_1 - \lambda_2|}\varphi_D(x,y). \quad (12)$$

In other words, $\varphi_S(x,y)$ is $(\lambda_2 + 3\lambda_1)/|\lambda_1 - \lambda_2|$ times more sensitive than $\varphi_D(x,y)$. For the numerical case of $\lambda_1 = 632.8$ nm and $\lambda_2 = 532$ nm, the phase sensitivity gain is $G=24.11$. This means that the infrared-equivalent wavelength is $\lambda_D = 3,340$ nm while the equivalent wavelength for $\varphi_S(x,y)$ is $\lambda_S = 138.5$ nm, this is within the far-ultraviolet (FUV) range. This FUV-equivalent interferometric technique may be extended to even higher super-resolution interferometric measurements. For example demodulating a temporal interferogram set with $\lambda_1 = 632.8$ nm and seven interferogram sets with $\lambda_2 = 532$ nm, the equivalent-wavelength of $\varphi_S(x,y) = \varphi_1(x,y) + 7\varphi_2(x,y)$ is $\lambda_S = \lambda_1\lambda_2/(7\lambda_1 + \lambda_2) = 67$ nm, well into the extreme-ultraviolet (EUV) wavelength range.

### 4. Temporal 2-steps phase unwrapping applied to the FUV-equivalent phase

In this paper, the non-wrapped infrared-equivalent $\varphi_D(x,y) \in (0, 2\pi)$ is used as first estimation to temporarily unwrap FUV-equivalent $\varphi_S^W(x,y) = W[\varphi_S(x,y)]$ in just two steps; being $W[x] = angle[\exp(i\,x)]$. The extended-range 2-steps temporal phase-unwrapping formula is [13, 14],

$$\varphi_S(x,y) = \frac{\lambda_D}{\lambda_S}\varphi_D(x,y) + W\left[\varphi_S^W(x,y) - \frac{\lambda_D}{\lambda_S}\varphi_D(x,y)\right]. \quad (13)$$

The two conditions that need to be fulfilled for the above equation are,

$$\varphi_D(x,y) \in (0, 2\pi), \quad and \quad \left[\varphi_S(x,y) - \frac{\lambda_D}{\lambda_S}\varphi_D(x,y)\right] \in (0, 2\pi). \quad (14)$$

If these conditions are not fulfilled, the 2-steps temporal phase-unwrapping do not work [13].

An additional advantage of extended-range temporal unwrapping in Eq. (13) is its immunity to the phase errors (harmonics and noise) of $\varphi_D(x,y)$ [13]. This means that having the erroneous phases,

$$\begin{aligned}\varphi_D(x,y) &= \varphi_1(x,y) - \varphi_2(x,y) + e_D(x,y), \\ \varphi_S^W(x,y) &= W[\varphi_1(x,y) + 3\varphi_2(x,y) + e_S(x,y)].\end{aligned} \quad (15)$$

Being $e_D(x,y)$ and $e_S(x,y)$ degrading harmonics plus noise. The unwrapped phase $\varphi_S(x,y)$ in Eq. (13) contains only its own degrading signal $e_S(x,y)$,

$$\varphi_S(x,y) = \varphi_1(x,y) + 3\varphi_2(x,y) + e_S(x,y). \quad (16)$$

In other words, the phase error $e_D(x,y)$ in $\varphi_D(x,y)$ do not propagate towards the unwrapped phase $\varphi_S(x,y)$. This was mathematically proven in [13].

### 5. Signal-to-noise power-ratio between the phase-sum and phase-difference

As previously seen the demodulated phases $\varphi_1(x,y)$, $\varphi_2(x,y)$, $\varphi_3(x,y)$ and $\varphi_4(x,y)$ are always corrupted by noise. Therefore the demodulated phases are,

$$\begin{aligned}
\varphi_1(x,y) &= (2\pi/\lambda_1)W(x,y) + n_1(x,y); \\
\varphi_2(x,y) &= (2\pi/\lambda_2)W(x,y) + n_2(x,y); \\
\varphi_3(x,y) &= (2\pi/\lambda_2)W(x,y) + n_3(x,y); \\
\varphi_4(x,y) &= (2\pi/\lambda_2)W(x,y) + n_4(x,y).
\end{aligned} \qquad (17)$$

As said $\{n_1, n_2, n_3, n_4\}$ are independent Gaussian zero-mean stationary random processes [15]. Therefore these noise fields have the same average $\mu$ and the same standard deviation $\sigma^2$,

$$\begin{aligned}
\mu &= E\{n_1(x,y)\} = E\{n_2(x,y)\} = E\{n_3(x,y)\} = E\{n_4(x,y)\}; \\
\sigma^2 &= E\{n_1^2(x,y)\} = E\{n_2^2(x,y)\} = E\{n_3^2(x,y)\} = E\{n_4^2(x,y)\}.
\end{aligned} \qquad (18)$$

Where $E\{\cdot\}$ is the ensemble average of its argument [15]. Also it is assumed that the noise amplitudes are small *i.e.* $\sigma \ll 2\pi$. Using Eq. (17), the phase-difference and phase-sum are,

$$\begin{aligned}
\varphi_D(x,y) &= \frac{2\pi}{\lambda_D}W(x,y) + n_D(x,y); \\
\varphi_S(x,y) &= \frac{2\pi}{\lambda_S}W(x,y) + n_S(x,y).
\end{aligned} \qquad (19)$$

Where $n_D = n_1 - n_2$ and $n_S = n_1 + n_2 + n_3 + n_4$. Then the signal-to-noise power-ratio for $\varphi_D(x,y)$ and $\varphi_S(x,y)$ are,

$$\begin{aligned}
\left(\frac{S}{N}\right)_{\varphi_D(x,y)} &= \frac{\iint_{(x,y)\in\Omega} |\varphi_D|^2 \, d\Omega}{\iint_{(x,y)\in\Omega} |n_D|^2 \, d\Omega} = \frac{\left(\frac{2\pi}{\lambda_D}\right)^2 \iint_{(x,y)\in\Omega} |W|^2 \, d\Omega}{\iint_{(x,y)\in\Omega} |n_1 - n_2|^2 \, d\Omega}; \\
\left(\frac{S}{N}\right)_{\varphi_S(x,y)} &= \frac{\iint_{(x,y)\in\Omega} |\varphi_S|^2 \, d\Omega}{\iint_{(x,y)\in\Omega} |n_S|^2 \, d\Omega} = \frac{\left(\frac{2\pi}{\lambda_S}\right)^2 \iint_{(x,y)\in\Omega} |W|^2 \, d\Omega}{\iint_{(x,y)\in\Omega} |n_1 + n_2 + n_3 + n_4|^2 \, d\Omega}.
\end{aligned} \qquad (20)$$

Being $\Omega$ the region of well-defined fringe-data and the coordinates $(x,y)$ where deleted for the sake of clarity. Given that the noise samples $\{n_1, n_2, n_3, n_4\}$ are generated by the same stationary random process, *in the average*, the energy of $n_S = n_1 + n_2 + n_3 + n_4$ is four times higher than $n_D = n_1 - n_2$ [15],

$$\iint_{(x,y)\in\Omega} |n_2 + n_2 + n_3 + n_4|^2 \, d\Omega = 4 \iint_{(x,y)\in\Omega} |n_1 - n_2|^2 \, d\Omega. \qquad (21)$$

Therefore, the signal-to-noise power-ratio gain between $\varphi_S(x,y)$ and $\varphi_D(x,y)$ is,

$$\frac{(S/N)_{\varphi_S(x,y)}}{(S/N)_{\varphi_D(x,y)}} = \frac{1}{4}\left(\frac{\lambda_D}{\lambda_S}\right)^2 = \frac{1}{4}\left(\frac{\lambda_2 + 3\lambda_1}{\lambda_2 - \lambda_1}\right)^2. \qquad (22)$$

The super-resolved, FUV-equivalent phase $\varphi_S(x,y)$ has $0.25[(\lambda_2 + 3\lambda_1)/(\lambda_2 - \lambda_1)]^2$ higher signal-to-noise power-ratio than $\varphi_D(x,y)$. For the numerical case analyzed with wavelengths $\lambda_1 = 632.8\text{nm}$ and $\lambda_2 = 532\text{nm}$, the FUV-equivalent phase-sum $\varphi_S(x,y)$ has 145 times higher signal-to-noise ratio than $\varphi_D(x,y)$. Therefore, not only $\varphi_S(x,y)$ is more sensitive ($\lambda_S = 138.5\text{nm}$), but also it has higher signal-to-noise power-ratio than $\varphi_D(x,y)$.

## 6. Summary

We have presented a far-ultraviolet (FUV) equivalent-wavelength super-resolution interferometry technique. This FUV-equivalent interferometry method uses four visible-wavelength phase estimations $\{\varphi_1(x,y), \varphi_2(x,y), \varphi_3(x,y), \varphi_4(x,y)\}$ (see Eq. (4)). We then compute the phase-difference $\varphi_D = \varphi_1 - \varphi_2$ and the phase-sum $\varphi_S = \varphi_1 + \varphi_2 + \varphi_3 + \varphi_4$. The phase-sensitivity between $\varphi_1(x,y)$ and $\varphi_2(x,y)$ is made close enough so their difference lie within the infrared sensitivity range and non-wrapped i.e. $\varphi_D(x,y) \in (0, 2\pi)$. In contrast the phase-sum $\varphi_S(x,y)$ lie within the FUV sensitivity-range and it is highly wrapped. Finally we used an extended-range 2-steps temporal unwrapper to unwrap $\varphi_S(x,y)$ taking $\varphi_D(x,y)$ as first estimation [13, 14]. As far as we know this is the first FUV-equivalent phase-sensitivity super-resolution interferometric technique ever reported.

Now we want to clearly point-out the main contributions of this paper as follows,

a) Previous dual-frequency close-sensitivity extended-range interferometry techniques used just the phase-difference of two sets of fringe-patterns [3-20].
b) Here we obtain four phase estimations $\{\varphi_1(x,y), \varphi_2(x,y), \varphi_3(x,y), \varphi_4(x,y)\}$ (Eq. (4)) from four sets of temporal visible-wavelength interferograms.
c) The infrared-equivalent phase-difference $\varphi_D(x,y) = \varphi_1(x,y) - \varphi_2(x,y)$ would suffice to analyze an extended-range phase measurement, because this phase is already unwrapped $\varphi_D(x,y) \in (0, 2\pi)$. However as we saw, this phase-difference is much noisier than the FUV-equivalent phase-sum $\varphi_S(x,y)$.
d) In contrast the FUV equivalent-wavelength of $\varphi_S = \varphi_1 + \varphi_2 + \varphi_3 + \varphi_4$ has much higher phase-sensitivity and signal-to-noise ratio but it is highly wrapped. Spatial super-resolution phase-unwrapping of discontinuous optical wavefronts coded in $\varphi_S(x,y)$ is precluded.
e) Therefore to unwrap the higher quality phase-sum $\varphi_S(x,y)$ we take the phase-difference $\varphi_D(x,y)$ as first estimation using our previously published 2-steps extended-range temporal phase unwrapper [13, 14]
f) As far as we know, no other super-resolution FUV equivalent-wavelength interferometric phase demodulating method has been proposed to this day.

Of course this super-resolution interferometric technique may be extended to consider more independently demodulated phases. For example, phase demodulating one interferogram with a wavelength of $\lambda_1 = 632.8\text{nm}$ and seven interferograms with wavelength $\lambda_2 = 532\text{nm}$, then the equivalent wavelength-sensitivity of the phase sum $\varphi_S(x,y) = \varphi_1(x,y) + 7\varphi_2(x,y)$ will be $\lambda_S = \lambda_1 \lambda_2 / (7\lambda_1 + \lambda_2) = 67\text{nm}$, well into the extreme-ultraviolet (EUV) wavelength range.